\documentclass[italian,english]{article}
\usepackage[latin9]{inputenc}
\usepackage{color}
\usepackage{amssymb}

\newcommand{\lyxaddress}[1]{
\par {\raggedright #1
\vspace{1.4em}
\noindent\par}
}

\usepackage{babel}

\begin{document}

\title{\textbf{Gravitational waves in the Hyperspace?}}

\author{\textbf{Christian Corda$^{*}$, Giorgio Fontana$^{+}$ and Gloria
Garcia Cuadrado$^{\&}$}}

\maketitle

\lyxaddress{\begin{center}
$^{*}$Associazione Scientifica Galileo Galilei, Via Pier Cironi
16 - 59100 Prato, Italy and 0574news.it - Sezione Scientifico-Tecnologica,
via Sante Pisani 46 - 59100 Prato, Italy; $^{+}$Università di Trento,
38050 Povo (Trento), Italy; $^{\&}$Aerospace Tecnology Center, Av.
Segle XXI - E-08840 Viladecans (Barcelona), Spain
\par\end{center}}

\lyxaddress{\begin{center}
\textit{E-mail addresses:} \textcolor{blue}{$^{*}$christian.corda@ego-gw.it;
$^{+}$giorgio.fontana@unitn.it; gloria.garcia@ctae.org}
\par\end{center}}
\begin{abstract}
In the framework of the debate on high-frequency gravitational waves
(GWs), after a review of GWs in standard General Relativity, which
is due for completness, the possibility of merging such a traditional
analysis with the Hyperspace formalism that has been recently introduced
in some papers in the literature, with the goal of a better understanding
of manifolds dimensionality also in a cosmological framework, is discussed.
Using the concept of refractive index in the Hyperspace, spherical
solutions are given and the propagation of GWs in a region of the
Hyperspace with an unitary refractive index is also discussed. Propagation
phenomena associated to the higher dimensionality are proposed, possibly
including non-linear effects. Further and accurate studies in this
direction are needed.\end{abstract}
\begin{itemize}
\item PACS numbers: 04.80.Nn, 04.30.Nk, 04.30.-w
\item Keywords: gravitational waves; General Relativity; Hyperspace; Cosmology.
\end{itemize}
The data analysis of interferometric GWs detectors has recently been
started (for the current status of GWs interferometers see \cite{key-1,key-2,key-3,key-4,key-5,key-6,key-7,key-8})
and the scientific community aims at a first direct detection of GWs
in next years. 

Detectors for GWs will be important for a better knowledge of the
Universe and either to confirm or ruling out the physical consistency
of General Relativity or any other theory of gravitation \cite{key-9,key-10,key-11,key-12,key-13,key-14}.
In fact, in the context of Extended Theories of Gravity, some differences
between General Relativity and the other theories can be pointed out
starting from the linearized theory of gravity \cite{key-9,key-10,key-12,key-14}.

Recently, some papers in the literature have shown the importance
of high-frequency GWs \cite{key-15,key-16,key-17}. In this context,
a difference approach to gravity has also been proposed with the Hyperspace
formalism, performing a coordinate-transformation from spacetime to
Euclidean coordinates \cite{key-18,key-19}. 

After a review of GWs in standard General Relativity, which is due
for completness, in this letter we discuss the possibility of merging
such a traditional analysis with the Hyperspace formalism that has
been recently introduced in some papers in the literature with the
goal of a better understanding of manifolds dimensionality also in
a cosmological framework \cite{key-18,key-19,key-20}. Using the concept
of refractive index in the Hyperspace, we find spherical solutions
and we also discuss the propagation of GWs in a region of the Hyperspace
with an unitary refractive index. 

After this, propagation phenomena associated to the higher dimensionality
are proposed, possibly including non-linear effects. Further and accurate
studies in this direction are needed.

Gravitational waves in General Relativity\textbf{ }have been analyzed
in lots of works in literature, starting by the work of the Bondi's
research group \cite{key-20} expecially in the Transverse-Traceless
(TT) gauge \cite{key-21,key-22,key-23}. In the first part of this
letter, we reanalyze the TT gauge for GWs following \cite{key-28}.

In the context of General Relativity, working with $c=1$ and $\hbar=1$,
calling $\widetilde{R}_{\mu\nu\rho\sigma}$ , $\widetilde{R}_{\mu\nu}$
and $\widetilde{R}$ the linearized quantity which correspond to $R_{\mu\nu\rho\sigma}$
, $R_{\mu\nu}$ and $R$, and putting \begin{equation}
g_{\mu\nu}=\eta_{\mu\nu}+h_{\mu\nu}\textrm{ with }\mid h_{\mu\nu}\mid\ll1\label{eq: linearizza}\end{equation}
the linearized Einstein field equations can be written like \cite{key-22,key-24} 

\begin{equation}
\widetilde{R}_{\mu\nu}-\frac{\widetilde{R}}{2}\eta_{\mu\nu}=0.\label{eq: linearizzate1}\end{equation}

Defining 

\begin{equation}
\bar{h}_{\mu\nu}\equiv h_{\mu\nu}-\frac{h}{2}\eta_{\mu\nu}\label{eq: h barra}\end{equation}

and putting eq. (\ref{eq: h barra}) in eqs. (\ref{eq: linearizzate1})
it is

\begin{equation}
\bar{\square h}_{\mu\nu}-\partial_{\mu}(\partial^{\alpha}\bar{h}_{\alpha\nu})-\partial_{\nu}(\partial^{\alpha}\bar{h}_{\alpha\mu})+\eta_{\mu\nu}\partial^{\beta}(\partial^{\alpha}\bar{h}_{\alpha\beta}),\label{eq: onda}\end{equation}

where $\square$ is the D'Alembertian operator.

Now, let us consider the gauge transform (Lorenz condition)

\begin{equation}
\bar{h}_{\mu\nu}\rightarrow\bar{h}'_{\mu\nu}=\bar{h}_{\mu\nu}-\partial_{(\mu}\epsilon_{\nu)}+\eta_{\mu\nu}\partial^{\alpha}\epsilon_{\alpha},\label{eq: gauge lorenzt}\end{equation}

with the condition $\square\epsilon_{\nu}=\partial^{\mu}\bar{h}_{\mu\nu}$
for the parameter $\epsilon^{\mu}$. It is

\begin{equation}
\partial^{\mu}\bar{h}'_{\mu\nu}=0,\label{eq: cond lorentz}\end{equation}

and, omitting the $'$, the field equations can be rewritten like

\begin{equation}
\square\bar{h}_{\mu\nu}=0.\label{eq: onda T}\end{equation}

Solutions of eqs. (\ref{eq: onda T}) are plane waves, i.e.

\begin{equation}
\bar{h}_{\mu\nu}=A_{\mu\nu}(\overrightarrow{k})\exp(ik^{\alpha}x_{\alpha})+c.c.\label{eq: sol T}\end{equation}

which take the conditions

\begin{equation}
\begin{array}{c}
k^{\alpha}k_{\alpha}=0\\
\\k^{\mu}A_{\mu\nu}=0,\end{array}\label{eq: vincoli}\end{equation}

which arises respectively from the linearized field equations and
from eq. (\ref{eq: cond lorentz}). The first of eqs. (\ref{eq: vincoli})
shows that perturbations have the speed of the light, the second represents
the transverse property of the waves.

Fixed the Lorenz gauge, another transformation with $\square\epsilon^{\mu}=0$
can be performed; one takes

\begin{equation}
\begin{array}{c}
\square\epsilon^{\mu}=0\\
\\\partial_{\mu}\epsilon^{\mu}=0,\end{array}\label{eq: gauge2}\end{equation}

which works because $\square\bar{h}=0$. Then

\begin{equation}
\begin{array}{ccc}
\bar{h}=0 & \Rightarrow & \bar{h}_{\mu\nu}=h_{\mu\nu},\end{array}\label{eq: h ug h}\end{equation}

(traceless property) i.e. $h_{\mu\nu}$ is a transverse plane wave
too. The gauge transformation (\ref{eq: gauge2}) also saves the conditions

\begin{equation}
\begin{array}{c}
\partial^{\mu}\bar{h}_{\mu\nu}=0\\
\\\bar{h}=0.\end{array}\label{eq: vincoli 2}\end{equation}

Considering a wave incoming in the positive $z$ direction it is

\begin{equation}
k^{\mu}=(k,0,0k)\label{eq: k}\end{equation}

and the second of eqs. (\ref{eq: vincoli}) implies

\begin{equation}
\begin{array}{c}
A_{0\nu}=-A_{3\nu}\\
\\A_{\nu0}=-A_{\nu3}\\
\\A_{00}=-A_{30}+A_{33}.\end{array}\label{eq: A}\end{equation}

Now, one has to compute the freedom degrees of $A_{\mu\nu}$. As $A_{\mu\nu}$
is a symmetric tensor, 10 components were present at the begining
of the analysis. 3 components have been lost for the transverse property,
more, the condition (\ref{eq: h ug h}) reduces the components to
6. One can take $A_{00}$, $A_{11}$, $A_{22}$, $A_{21}$, $A_{31}$,
$A_{32}$ like independent components; another gauge freedom can be
used to put to zero three more components (i.e. one can only chose
three of $\epsilon^{\mu}$, the fourth component depends from the
others by $\partial_{\mu}\epsilon^{\mu}=0$).

Then, taking

\begin{equation}
\begin{array}{c}
\epsilon_{\mu}=\tilde{\epsilon}_{\mu}(\overrightarrow{k})\exp(ik^{\alpha}x_{\alpha})+c.c.\\
\\k^{\mu}\tilde{\epsilon}_{\mu}=0,\end{array}\label{eq: ancora gauge}\end{equation}

the transform law for $A_{\mu\nu}$ is (see eqs. (\ref{eq: gauge lorenzt})
and (\ref{eq: sol T}) )

\begin{equation}
A_{\mu\nu}\rightarrow A'_{\mu\nu}=A_{\mu\nu}-2ik(_{\mu}\tilde{\epsilon}_{\nu}).\label{eq: trasf. tens.}\end{equation}

As for the six components of interest only three are gauge-invariants,
$\tilde{\epsilon}_{\nu}$ can be chosen to put equal to zero the others
(see \cite{key-28} for details). From the traceless property two
components are also equal \cite{key-28}. In this way, only two physical
components are present and the total perturbation of a gravitational
wave propagating in the $z+$ direction in this gauge is \cite{key-28}

\begin{equation}
h_{\mu\nu}(t-z)=A^{+}(t-z)e_{\mu\nu}^{(+)}+A^{\times}(t-z)e_{\mu\nu}^{(\times)},\label{eq: perturbazione totale}\end{equation}

that describes the two polarizations of gravitational waves which
arises from General Relativity in the transverse-traceless (TT) gauge
(ref. \cite{key-20,key-21,key-22}). This gauge is historically called
TT, because in these particular coordinates the gravitational waves
have a transverse effect and are traceless.

A different approach on gravity has been recently proposed considering
the Hyperspace formalism, performing a coordinate-transformation from
spacetime to Euclidean coordinates \cite{key-18,key-19}. In this
framework, a different space-time structure is introduced \cite{key-16,key-18,key-19,key-25,key-26}.
One starts by the flat space-time line element 

\begin{equation}
d\tau^{2}=-dt^{2}+dz^{2}+dx^{2}+dy^{2},\label{eq: metrica piatta}\end{equation}
and re-arranges like \cite{key-16,key-18,key-19,key-25,key-26}

\begin{equation}
dt^{2}=d\tau^{2}+dz^{2}+dx^{2}+dy^{2}.\label{eq: smetrica piatta}\end{equation}

Gravity is considered by resorting the analogy to optical propagation
in the 3-space. In this way, a gravitational refractive index of the
Hyperspace named $n$ is introduced \cite{key-25,key-27}

\begin{equation}
dt^{2}=n^{2}(d\tau^{2}+dz^{2}+dx^{2}+dy^{2}).\label{eq: smetrica piatta n}\end{equation}

The case $n=1$ gives the Lorentzian flat space-time (\ref{eq: metrica piatta}),
while, switching to spherical coordinates and putting $d\theta=d\varphi$,
the condition \cite{key-25,key-27}

\begin{equation}
n=(1-\frac{2\tilde{G}m}{r})^{-1}\label{eq: n}\end{equation}

generates the well known Schwarzschild space-time \cite{key-22}.

One can also introduce directional indexes of the Hyperspace

\begin{equation}
dt^{2}=n_{\tau}^{2}d\tau^{2}+n_{z}^{2}dz^{2}+n_{x}^{2}dx^{2}+n_{y}^{2}dy^{2}),\label{eq: smetrica piatta 4n}\end{equation}

and, recalling that the GWs line element which arises by eq. (\ref{eq: perturbazione totale}),
considering only the {}``$+$'' polarization, is \cite{key-22,key-28}

\begin{equation}
d\tau^{2}=-dt^{2}+dz^{2}+[1+A^{+}(t-z)]dx^{2}+[1-A^{+}(t-z)]dy^{2},\label{eq: metrica +}\end{equation}
one obtains 

\begin{equation}
\begin{array}{c}
n_{\tau}^{2}=1\\
\\n_{z}^{2}=1\\
\\n_{x}^{2}=1+A^{+}(t-z)\\
\\n_{y}^{2}=1-A^{+}(t-z).\end{array}\label{eq: enni}\end{equation}

More in general, assuming an isotropic propagation, one consider a
generic wave equation for weak waves travelling at the speed of light
\cite{key-25}

\begin{equation}
\square n=0.\label{eq: onda n}\end{equation}

In Hyperspace coordinates, by separing variables, one looks for solutions
of the type $n(r,t)=R(r)T(T)$, obtaining

\begin{equation}
\begin{array}{c}
r\frac{d^{2}R}{dr^{2}}+3\frac{dR}{dr}+k^{2}rR=0\\
\\\frac{d^{2}T}{dt^{2}}+\omega^{2}T=0.\end{array}\label{eq: onde sferiche}\end{equation}

In the flat background the refractive index is unitary. To study non
linear effects one has to include such a value in the expression of
the speed. In this way eq. (\ref{eq: onda n}) changes into 

\begin{equation}
(n+1)^{2}\frac{\Game^{2}n}{\Game t^{2}}=\triangledown^{2}n=0,\label{eq: onda n quadra}\end{equation}

which can be re-written like

\begin{equation}
\square n=-(n^{2}+2n)\frac{\Game^{2}n}{\Game t^{2}}.\label{eq: onda n quadra 2}\end{equation}

Eq. (\ref{eq: onda n}) is re-obtained in the limit $n\ll1.$

These equations show that if a beam of GWs is more intense at center
than at wings, the central part travels at lower four-speed (self-focusing,
see also the classical tratment in \cite{key-29}).

The unitary background refractive index is attributed to a cosmological
GWs background which supports the spacetime \cite{key-25}. The existence
of such a cosmological GWs background has been postulated by lots
works in the literature, see for example \cite{key-30,key-31,key-32,key-33}
and it has been recently analysed in the framework of Extended Theories
of Gravity \cite{key-34,key-35,key-36}.

Removing the background, the non-linear eq. (\ref{eq: onda n quadra})
becomes \begin{equation}
n^{2}\frac{\Game^{2}n}{\Game t^{2}}=\triangledown^{2}n=0.\label{eq: onda n quadra 3}\end{equation}

In this case, solutions of the type $n(r,t)=R(r)T(T)$ give \cite{key-25}
\begin{equation}
\begin{array}{c}
r\frac{d^{2}R}{dr^{2}}+3\frac{dR}{dr}-\varkappa rR^{2}=0\\
\\\frac{d^{2}T}{dt^{2}}+\frac{-\varkappa}{T}=0.\end{array}\label{eq: onde sferiche 2}\end{equation}

The second of eqs. (\ref{eq: onde sferiche 2}) is a non linear differential
equation of the Emden-Fowler type \cite{key-37}, which can be integrated
only numerically in terms of $T(t)$ if $\varkappa\neq0.$ It is possible
to write 

\begin{equation}
t=C_{1}+\int dt(2\varkappa lnT)^{-\frac{1}{2}}\label{eq: t}\end{equation}

where $C_{1}$ and $C_{2}$ are two real integration constants which
work in the numerical integration

\begin{equation}
T(t)\approx C_{1}+C_{2}t^{\alpha}\label{eq: T}\end{equation}

and $\alpha\approx1.03.$ Note: solution with both of $C_{1}\neq0$
and $C_{2}=0$ are not acceptable as $T(t)=constant$ is not solution
of the second of eqs. (\ref{eq: onde sferiche 2}). In such a numerical
integration the slope and the sign are given by initial conditions
and the equation is singular in the case $C_{1}=0.$

With $\varkappa=0$ one gets $T(t)=C_{1}+C_{2}t,$ while with $\varkappa<0$
the solution of the first of eqs. (\ref{eq: onde sferiche 2}) is
a damped oscillator. 

A possible interpretation of the results \cite{key-25} is that the
large scale sweeping of spacetimes through the Hyperspace becomes
slower and slower with time, while, because of the increasimg temporal
component of $n,$ the oscillating and decreasing spatial component
of $n,$ combined with a specific spacetime motion results in a nearly
constant $n$ for the local background, during the spacetime evolution
\cite{key-25}. The $r$-dependence in the first of eqs. (\ref{eq: onde sferiche 2})
implies that spacetime regions in which the particles propagation
speed can be different from the speed of the light, and the presence
of Hyperspace {}``highways'' is not totally excluded. Further and
accurate studies are needed in this direction.

In the Hyperspace $n$ can also oscillate from positive to negative
values, and further studies will be needed to understand the effect
of the sign change too. 

$\varkappa=0$ gives also $R(r)=-3\ln r+C_{3}$ \cite{key-25}, thus,
in this particular case the solution of eqs. (\ref{eq: onde sferiche 2})
is

\begin{equation}
n(r,t)=(C_{1}+C_{2}t)(-3\ln r+C_{3}).\label{eq: sol. part.}\end{equation}

Again, this particular solution shows that the speed of spacetimes
changes in the Hyperspace because both of evolution in time and expansion
in space.

\subsubsection*{Conclusion remarks}

In the framework of the debait on high-frequency GWs, after a review
of GWs in standard General Relativity, which is due for completness,
the possibility of merging such a traditional analysis with the Hyperspace
formalism which has been recently introduced in some papers in the
literature, with the goal of a better understanding of manifolds dimensionality
also in a cosmological framework, has been discussed. Using the concept
of refractive index in the Hyperspace, spherical solutions have been
given and the propagation of GWs in a region of the Hyperspace with
an unitary refractive index has been also discussed. Propagation phenomena
associated to the higher dimensionality have been proposed, possibly
including non-linear effects. Further and accurate studies in this
direction are needed.

\subsubsection*{Aknlowdgements}

This letter has been partially supported by the Sezione Scientifico-Tecnologica
of 0574news.it, via Sante Pisani 46, 59100 Prato, Italy


\begin{thebibliography}{37}
\bibitem{key-1}F. Acernese et cal. (the Virgo Collaboration) - Class.
Quant. Grav. \textbf{24,} 19, S381-S388 (2007)

\bibitem{key-2}C. Corda - Astropart. Phys. \textbf{27,} No 6, 539-549
(2007)

\bibitem{key-3}C. Corda - Int. J. Mod. Phys. D \textbf{16,} 9, 1497-1517
 (2007) 

\bibitem{key-4}B. Willke et al. - Class. Quant. Grav. \textbf{23}
8S207-S214 (2006) 

\bibitem{key-5}D. Sigg (for the LIGO Scientific Collaboration) -
www.ligo.org/pdf\_public/P050036.pdf

\bibitem{key-6}B. Abbott et al. (the LIGO Scientific Collaboration)
- Phys. Rev. D 72, 042002 (2005) 

\bibitem{key-7}M. Ando and the TAMA Collaboration - Class. Quant.
Grav. \textbf{19} 7 1615-1621 (2002)

\bibitem{key-8}D. Tatsumi, Y. Tsunesada and the TAMA Collaboration
- Class. Quant. Grav. \textbf{21} 5 S451-S456 (2004) 

\bibitem{key-9}S. Capozziello - \textit{Newtonian Limit of Extended
Theories of Gravity} in \textit{Quantum Gravity Research Trends} Ed.
A. Reimer, pp. 227-276 Nova Science Publishers Inc., NY (2005) - \foreignlanguage{italian}{also
in arXiv:}gr-qc/0412088 (2004);E. Elizalde, S. Nojiri, and S.D. Odintsov
- Phys. Rev. D 70, 043539 (2004) 

\bibitem{key-10}S. Capozziello and A. Troisi - Phys. Rev. D \textbf{72}
044022 (2005);G. Cognola, E. Elizalde, S. Nojiri, S.D. Odintsov and
S. Zerbini - J. Cosmol. Astropart. Phys. JCAP0502(2005)010

\bibitem{key-11}G. Allemandi, M. Capone, S. Capozziello and M. Francaviglia
- Gen. Rev. Grav. 38 1 (2006); G. Allemandi, M. Francaviglia, M. L.
Ruggiero and A . Tartaglia - Gen. Rel. Grav. 37 11 (2005) 

\bibitem{key-12}S. Capozziello and C. Corda - Int. J. Mod. Phys.
D \textbf{15} 1119 -1150 (2006);S. Nojiri and S.D. Odintsov - hep-th
0601213 (2006)

\bibitem{key-13}G. Cognola, E. Elizalde, S. Nojiri, S.D. Odintsov
and S. Zerbini - Phys. Rev. D 73, 084007 (2006)

\bibitem{key-14}C. Corda- Astropart. Phys. 28, 247-250 (2007);G.
Cognola, E. Elizalde, S. Nojiri, S.D. Odintsov, L. Sebastiani, S.
Zerbini - Phys. Rev. D 77, 046009 (2008)

\bibitem{key-15}F. Li, R.M.L. Baker Jr., Z. Fang, G. V. Stephenson
and Z. Chen - arXiv:0806.1989, accepted for publication by Eur. Phys.
Journ. C

\bibitem{key-16}G. Fontana, P. Murad and R.M.L. Baker Jr. - AIP Conference
Proceedings 880, 1117-1124, Melville, New York (2007)

\bibitem{key-17}P. Murad and R.M.L. Baker Jr. - $39^{h}$ AIAA/ASME/SAE/ASEE
Propulsion Conference Proceedings N. 2003-4882, Hunsville, Alabama,
USA (2003)

\bibitem{key-18}J.M.C. Mountanus - Found. of Phys. 31, 1357-1359
(2001)

\bibitem[19]{key-19}A. Gersten - Found. of Phys. 33, 1237-1251 (2003)

\bibitem[20]{key-20}Bondi H, Pirani FAE and Robinson I - Proc. Roy.
Soc. Lond. \textbf{A251} 519-533 (1959)

\bibitem{key-21}Rakhmanov M - Phys. Rev. D \textbf{71} 084003 (2005)

\bibitem{key-22}\foreignlanguage{italian}{Misner CW, Thorne KS and
Wheeler JA - {}``Gravitation'' - W.H.Feeman and Company - 1973}

\bibitem{key-23}Maggiore M- Physics Reports \textbf{331} 283-367
(2000)

\bibitem[24]{key-24}Landau L and Lifsits E - {}``Teoria dei campi''
- Editori riuniti edition III (1999)

\bibitem[25]{key-25}G. Fontana - AIP Conference Proceedings 978-0-7354-0486,
Melville, New York (2008)

\bibitem{key-26}G. Fontana - AIP Conference Proceedings 746, 1403
- 1410, Melville, New York (2005)

\bibitem[27]{key-27}J. B. Almeida - gr-qc/040722

\bibitem[28]{key-28}C. Corda - Int. Journ. Mod. Phys. A 22, 26, 4859-4881
(2007)

\bibitem[29]{key-29}V. Ferrari - Phys. Rev. D 37, 10, 3061-3064 (1988)

\bibitem[30]{key-30}B. Allen - Proceedings of the Les Houches School
on Astrophysical Sources of Gravitational Waves, eds. Jean-Alain Marck
and Jean-Pierre Lasota (Cambridge University Press, Cambridge, England
1998).

\bibitem{key-31}B. Allen - Phys. Rev. D \foreignlanguage{italian}{\textbf{3}}\textbf{7},
2078 (1988)

\bibitem{key-32}\foreignlanguage{italian}{L. P. Grishchuk and others
- Phys. Usp. 44 1-51 (2001)}

\bibitem{key-33}\foreignlanguage{italian}{L. P. Grishchuk and others
- Usp. Fiz. Nauk 171 3 (2001)}

\selectlanguage{italian}%
\bibitem[34]{key-34}\foreignlanguage{english}{C. Corda, S. Capozziello
and M. F. De Laurentis - AIP Conference Proceedings, Volume 966, pp
257-263 (2007) - Proceedings of the Fourth Italian-Sino Workshop on
Relativistic Astrophysics July 20-30 2007, Pescara, Italy}

\selectlanguage{english}%
\bibitem[35]{key-35}S. Capozziello, C. Corda and M. F. De Laurentis
- Mod. Phys. Lett. A 22, 35, 2647-2655 (2007)

\bibitem[36]{key-36}C. Corda - Astropart. Phys., 30, 209-215 (2008)

\bibitem[37]{key-37}P. L. Sachdev - \emph{Nonlinear ordinary differential
equations and their applications}, Marcel Dekker, Inc., New York,
1991
\end{thebibliography}
\end{document}